\documentclass{article}
\usepackage{graphicx} % Required for inserting images
 \usepackage{amsmath}
 \usepackage{physics}
 \usepackage{bigints}
 \usepackage{jheppub}
 \usepackage{xcolor}
 \usepackage{comment}
 \usepackage{physics}

\newcommand{\f}{\varphi}
\newcommand{\gf}{\Gamma_{\f}}

\title{Reconstructing early universe evolution with gravitational waves from supercooled phase transitions}
\author{Adam Gonstal, Marek Lewicki and Bogumi{\l}a {\'S}wie{\.z}ewska}
\affiliation{
University of Warsaw, Faculty of Physics,\\
Pasteura 5, 02-093 Warsaw, Poland
}
\date{June 2024}
\emailAdd{adam.gonstal@fuw.edu.pl, marek.lewicki@fuw.edu.pl, bogumila.swiezewska@fuw.edu.pl}
\begin{document}

%~~~~~~~~~~~~~~~~~~~~~~~~~~~~~~~~~~~~~~~~~~~~~~~~~~~~~~~~~~~~~
\abstract{
We study gravitational waves from supercooled cosmological first-order phase transitions. If such a transition is followed by inefficient reheating, the evolution history of the universe is modified by a period of early matter domination. This leaves an imprint on the predicted gravitational-wave spectra. Using Fisher analysis we show the parameter space in reach of upcoming gravitational wave observatories where reheating can be probed due to its impact on the stochastic background produced by the transition. We use both the simplified geometric parametrisation and the thermodynamical one explicitly including the decay rate of the field undergoing the transition as a parameter determining the spectrum. We show the expansion history following the transition can be probed provided the transition is very strong which is naturally realised in classically scale invariant models generically predicting supercooling. Moreover, in such a scenario the decay rate of the scalar undergoing the phase transition, a parameter most likely inaccessible to accelerators, can be determined through the spectrum analysis.
}

\maketitle
%~~~~~~~~~~~~~~~~~~~~~~~~~~~~~~~~~~~~~~~~~~~~~~~~~~~~~~~~~~~~~

%~~~~~~~~~~~~~~~~~~~~~~~~~~~~~~~~~~~~~~~~~~~~~~~~~~~~~~~~~~~~~
\section{Introduction}
%~~~~~~~~~~~~~~~~~~~~~~~~~~~~~~~~~~~~~~~~~~~~~~~~~~~~~~~~~~~~~

Early-universe phenomena are so far poorly constrained due to the lack of data. Some \linebreak of them, such as inflation, topological defects and cosmological phase transitions related 
\linebreak to symmetry breaking, are expected to have left a potentially observable signature in the \linebreak form of stochastic gravitational-wave background (SGWB)~\cite{LISACosmologyWorkingGroup:2022jok}. Since the first detection of \linebreak gravitational waves (GW) by the LIGO/Virgo/KAGRA collaboration~\cite{LIGOScientific:2016aoc, LIGOScientific:2016sjg, LIGOScientific:2017bnn,LIGOScientific:2017vox,LIGOScientific:2017ycc, LIGOScientific:2018mvr, LIGOScientific:2020ibl, LIGOScientific:2021usb, KAGRA:2021vkt} almost a decade \linebreak ago, 
this new messenger has been proven to be within our reach.
Recently, the pulsar timing \linebreak array (PTA) collaborations~\cite{NANOGrav:2023gor, EPTA:2023fyk,Reardon:2023gzh} reported evidence of an SGWB. It can be explained by collisions of supermassive black holes taking place in the late universe~\cite{NANOGrav:2023hfp, EPTA:2023xxk, Ellis:2023oxs} but an explanation by a primordial source is also possible. In particular, a signal from a first-order phase transition (PT) fits the data well~\cite{Ratzinger:2020koh, NANOGrav:2021flc, Gouttenoire:2023bqy, Madge:2023dxc, Figueroa:2023zhu, Ellis:2023oxs, Lewicki:2024ghw, Lewicki:2024sfw, Goncalves:2025uwh, Li:2025nja, Costa:2025csj}. Since the observed signal of a large amplitude is in the nano-Hertz range, the corresponding PT sourcing such a signal would need to be extremely supercooled and the universe should not reheat too much afterwards. This poses a challenge in constructing realistic models featuring such a PT~\cite{Bai:2021ibt, Bringmann:2023opz}.

It is much more natural to expect a signal sourced by a PT in the LISA~\cite{LISA:2017pwj,Colpi:2024xhw} frequency range corresponding to processes at energies around the electroweak scale. In the case of a very high-temperature PT a signal visible in Einstein Telescope (ET)~\cite{Punturo:2010zz, Hild:2010id} could be observed. Supercooled PTs, in which the transition proceeds at temperatures much below the critical temperature with a huge latent heat release, are particularly promising in terms of the strength of the predicted GW signal~\cite{Randall:2006, Konstandin:2011, Jaeckel:2016, Hashino:2016, vonHarling:2017, Hambye:2018qjv, Baldes:2018emh, Prokopec:2018,Hashino:2018, Marfatia:2020, Bruggisser:2022rdm, Kierkla:2023von, Jinno:2016knw,Marzola:2017jzl,Marzo:2018nov,Fujikura:2019oyi,Mohamadnejad:2019vzg, Kang:2020jeg, Mohamadnejad:2021tke, Lewicki:2021xku,Schmitt:2024pby,Lewicki:2024sfw,Goncalves:2025uwh, Li:2025nja, Costa:2025csj}. Studying the observability of a PT-sourced GW signal at LISA and ET, and the analysis of reconstruction of the parameters of the spectrum as well as of the underlying fundamental physics model is the main goal of the present work. 

This issue has been studied in the past~\cite{Gowling:2021gcy, Giese:2021dnw, Boileau:2022ter, Gowling:2022pzb} for GWs sourced by bubble collisions in the plasma, and recently in ref.~\cite{Caprini:2024hue} also signals from bubble wall collisions, relevant for very strong transitions, have been included. In the present work, we focus on supercooled PTs, which can result in GWs produced via bubble collisions or sound waves in the plasma. Following ref.~\cite{Lewicki:2022pdb} we assume that the spectra generated via these two sources are identical for strongly supercooled transitions. We extend the results of ref.~\cite{Caprini:2024hue} for supercooled PTs studying the possibility of probing the modified expansion history of the universe via GWs. In particular, we are interested in probing a period of early matter domination, caused by inefficient reheating after a supercooled PT. This inefficiency is associated with a small decay rate of the field undergoing the transition to the SM fields and the resulting modification of the spectra is one of the very few ways in which features of the underlying particle physics models can be probed through GW signals.

The paper is organised as follows. In section~\ref{sec:PT} we introduce the physical setting under study, namely supercooled phase transition and the process of reheating. We then proceed with the description of the applied methods of analysis: in section~\ref{sec:templates} we provide the signal templates, in both geometrical (also called spectral) and thermodynamical parameterisations, in section~\ref{sec:noise} we review the models for detector noise applied, while section~\ref{sec:fisher} introduces the Fisher matrix method. The results of the parameter reconstruction are presented in section~\ref{sec:reconstruction-spectral} for spectral and in section~\ref{sec:reconstruction:thermo} for thermodynamical parameterisation. We conclude in section~\ref{sec:conclusions}.

%~~~~~~~~~~~~~~~~~~~~~~~~~~~~~~~~~~~~~~~~~~~~~~~~~~~~~~~~~~~~~
\section{Supercooled PT and modified evolution of the universe}
\label{sec:PT}
%~~~~~~~~~~~~~~~~~~~~~~~~~~~~~~~~~~~~~~~~~~~~~~~~~~~~~~~~~~~~~

This section briefly reviews the most important features of supercooled PTs and specifies our treatment of reheating after the PT.

A first-order phase transition is possible when the scalar effective potential features two minima separated by a barrier. At high temperatures, the symmetric minimum at the origin of the field space is the global one. As the universe cools down, the potential evolves and eventually the two minima become degenerate at the critical temperature, $T_c$. Just below this temperature, the phase transition is already possible since it is energetically favourable for the field to tunnel to the global minimum of the potential. 
If, however, the decay rate of the false vacuum is suppressed, the phase transition is delayed. At some temperature $T_V$ the energy stored in radiation becomes equal to the energy stored in the vacuum and a phase of thermal inflation begins. The phase transition begins when the nucleation rate is sufficient for at least one bubble of true vacuum to nucleate per Hubble volume. We consider the PT to be complete when the bubbles percolate at $T_*$. 

The strength of the transition for supercooled PTs can be approximated as~\cite{Ellis:2018mja,Ellis:2019oqb}
\begin{equation}
    \alpha=\frac{\Delta V}{\rho_{R*}},
\end{equation}
where $\Delta V$ denotes the energy difference between the two minima of the potential at the percolation temperature, which can be well approximated by $\Delta V$ evaluated at $T=0$. $\rho_{R*}$ denotes the energy density stored in the pre-existing radiation bath at percolation temperature. In general $\rho_R$ is given by
\begin{equation}
\label{eq:rho_rad}
    \rho_{R}(T)=\frac{\pi^2 g_*(T)}{30}T^4,
\end{equation}
with $g_*(T)$ counting the number of relativistic degrees of freedom. Typically, for supercooled PTs the pre-existing radiation is severely diluted and we can assume $\alpha \gg 1$. The inverse time scale of the transition is typically approximated by the first-order expansion of the false vacuum decay rate 
\begin{equation}
\label{eq:beta_Taylor}
\Gamma_V \propto   e^{\beta(t-t_*) + ...} \, .
\end{equation}
The decay rate can be computed in a given model using its effective potential~\cite{Linde:1980tt,Linde:1981zj}~\footnote{See~\cite{Ekstedt:2022tqk, Ekstedt:2023sqc,Gould:2024chm, Pirvu:2024nbe, Pirvu:2024ova, Chala:2024xll, Kierkla:2025qyz} for a recent discussion of this approximation.}. In order to remain as model-independent as possible we will not specify a potential and simply treat the inverse duration normalised to the Hubble rate $\beta/H$ as a free parameter.

The huge amount of energy stored in the potential and driving the thermal inflation is released after the PT as the latent heat of the transition. A fraction of this energy is converted to GWs while the rest remains in the scalar field until it is transferred to the primordial plasma in the process of reheating. There are two possible scenarios of reheating controlled by the relative magnitude of the decay rate of the scalar field that undergoes the transition, $\Gamma_{\f}$, to the Hubble rate at the moment of the transition, $H_*$. 

If $\gf\geqslant H_*$, the decay of the scalar field to plasma particles is fast, the reheating is efficient and can be considered instantaneous. Then, the reheating temperature can be easily estimated using energy conservation~\cite{Ellis:2019oqb} as follows. Since for strongly supercooled PTs $\alpha \gg 1$ we can assume that the total energy in the universe is dominated by $\Delta V$ at the moment of transition. This energy is to be transformed into the radiation energy of the reheated plasma at temperature $T_r$ so 
\begin{equation}
\label{eq:T-reh-instant}
    \Delta V= \rho_R (T_r),
\end{equation}
which leads to the approximation for the reheating temperature, $T_r\approx T_V$.

If $\gf < H_*$, the scalar cannot immediately decay into plasma particles and instead oscillates about the true minimum of the potential~\cite{Cutting:2018tjt,Cutting:2020nla}. The universe, when dominated by an oscillating scalar field, evolves as matter-dominated~\cite{Allahverdi:2020bys}~\footnote{In fact, the scalar field after the transition might not be exactly homogeneous. In such a case a modified equation of state would be needed. However, the exact result is not certain yet and we will continue to use the simplifying assumption of matter domination following an inefficient reheating.}. This causes a period of early matter domination that lasts until the decay rate $\gf$ becomes comparable with the Hubble parameter. The energy stored in the oscillating scalar field redshifts as matter so the final reheating temperature is lower than in the case of immediate reheating. To quantitatively study the reheating process we solve the following Boltzmann equations~\cite{Ellis:2020nnr} (see also refs.~\cite{Barenboim:2016mjm,Hambye:2018qjv,Pearce:2023kxp,Pearce:2025ywc} for a similar approach)
\begin{equation}
\label{eq:Boltzman}
\dot{\rho}_R+4 H \rho_R=\Gamma_{\varphi} \rho_{\varphi}, \quad \dot{\rho}_{\varphi}+3 H \rho_{\varphi}=-\Gamma_{\varphi} \rho_{\varphi}, \quad H^2=\frac{\rho_R+\rho_{\varphi}}{3 M_P^2},
\end{equation}
where the dot symbolises the time derivative, $\rho_R$ and $\rho_{\f}$ denote the energy density of radiation and the scalar field, respectively, and $M_P$ is the reduced Planck mass $M_P=2.435 \cdot 10^{18}$\ GeV. Since we assume that the transition is strongly supercooled, $\alpha\gg 1$, we set the initial energy density of the scalar field to $\rho_{\f}=\Delta V$, and the energy of radiation to zero.~\footnote{Numerically we set it to some very small number that does not influence the computations.} We account for the number of relativistic degrees of freedom changing with temperature $g_*(T)$ using data from ref.~\cite{Saikawa:2018rcs}. 
We solve this set of equations numerically until the field energy density becomes negligible~\footnote{By negligible we mean $\rho_\varphi/\rho_R=10^{-4}$~for LISA $\rho_\varphi/\rho_R=10^{-7}$ for ET. The different values have no physical importance and were chosen for increasing numerical accuracy.} and assume standard expansion from that point until today.

\begin{figure}[!ht]
    \centering
    \begin{minipage}{0.69\textwidth} % Width for the first figure
        \centering
        \includegraphics[width=\textwidth]{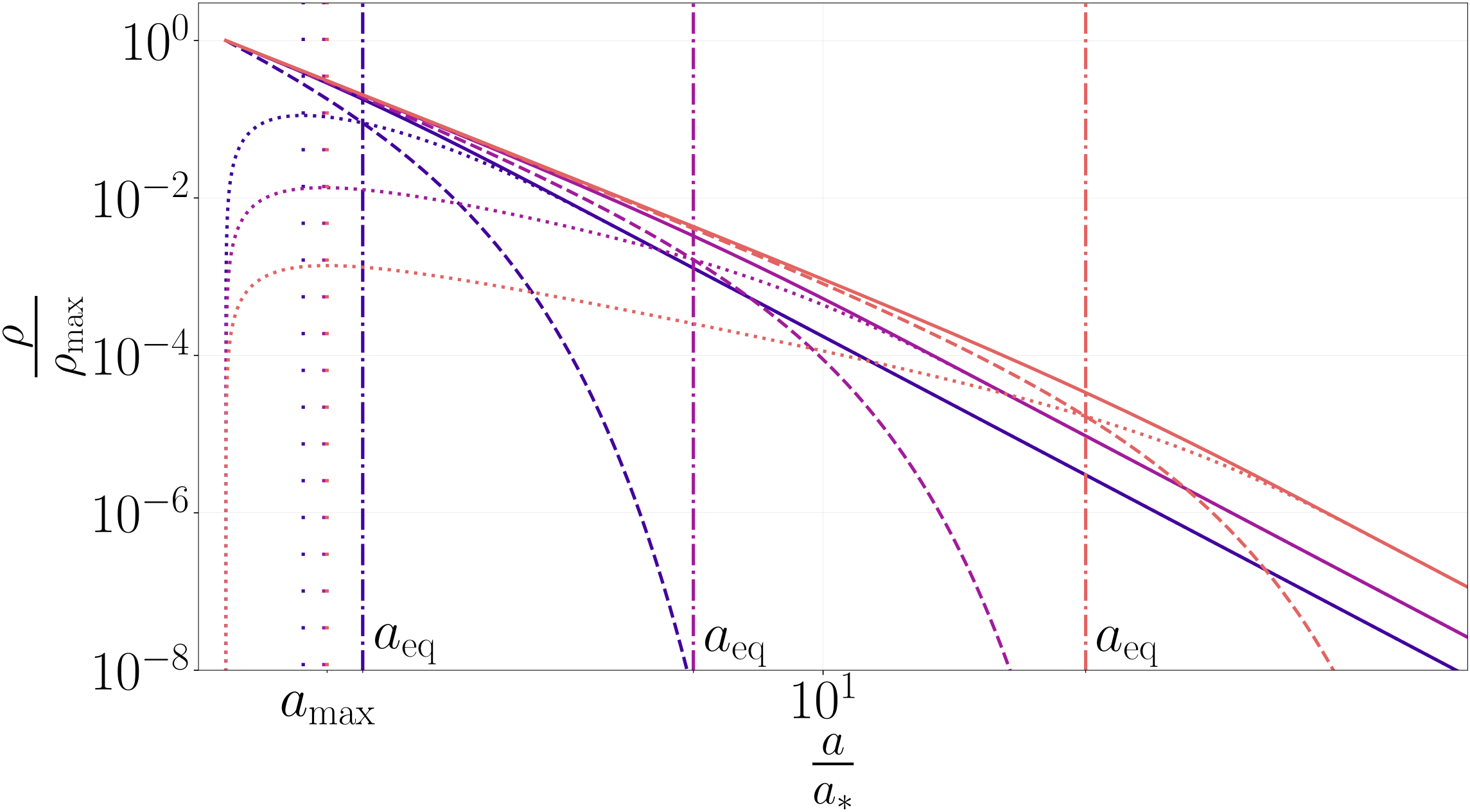}
    \end{minipage}
    \begin{minipage}{0.25\textwidth} % Width for the second figure
        \centering
        \includegraphics[width=\textwidth]{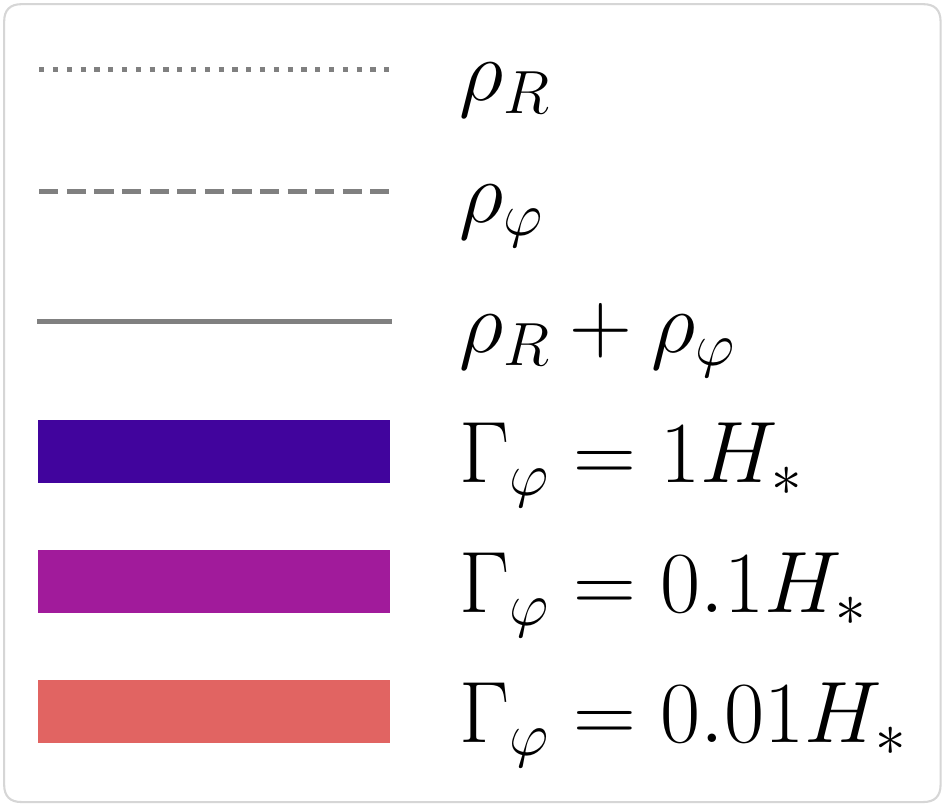}
    \end{minipage}        
    \caption{Evolution of the energy density of the scalar $\rho_\varphi$ (dashed lines) and the radiation $\rho_R$ (dotted lines) for different values of the scalar decay rate $\gf$ indicated with different colours. The densities are normalised to the overall energy scale $\Delta V=\rho_{\rm max}$. The vertical dotted lines to the left show the scale factor $a_{\rm max}$ for which the maximal temperature is reached,  while the vertical dot-dashed lines indicate the scale factor $a_{\rm eq}$ of reheating for which $\rho_\varphi=\rho_R$ in each case.
    }
    \label{fig:solutions-Boltzmann}
\end{figure}

The overall energy scale for the above set of equations is set by $\Delta V$, in particular, the Hubble parameter at the percolation temperature is given by $H_*^2=\Delta V/3M_P^2$. Therefore, to illustrate various reheating scenarios we fix $\Delta V$ and in figure~\ref{fig:solutions-Boltzmann} present the evolution of energy densities of radiation and the oscillating scalar field with the scale factor for different values of $\gf/H_*$, obtained by solving the system of eq.~\eqref{eq:Boltzman} numerically. %For $\gf/H_*=1$ the reheating is fast and the maximal temperature is close to the reheating temperature as predicted in eq.~\eqref{eq:T-reh-instant}. 
The maximal temperature of the plasma (obtained from $\rho_R$ using eq.~\eqref{eq:rho_rad}), $T_{\rm{max}}$, is reached for the scale factor $a_{\rm{max}}$, which weakly depends on $\Gamma_{\f}/H_*$. Yet, the value of $T_{\rm{max}}$ decreases with decreasing decay rate.  The scale of reheating, $a_{\rm{eq}}$, defined as the scale at which the energy densities of radiation and the scalar field are equal, grows with decreasing $\Gamma_{\f}/H_*$. Consequently, the corresponding reheating temperature $T_{\rm{eq}}$ decreases. 
To sum up, the lower the value of $\gf/H_*$, the longer the reheating phase and the lower the maximal temperature reached.

In the following sections, we use the maximal temperature, $T_{\rm{max}}$, to parameterise the GW spectra. We could equally choose the reheating temperature, $T_{\rm eq}$, since anyway the modified evolution history, as depicted in figure~\ref{fig:solutions-Boltzmann} is included in the computation of the redshift that GW signal produced during a first-order PT experiences. The resulting modification of the shape of the spectrum opens the possibility of probing the history of the universe using GWs from  PTs. Moreover, it could be possible to reconstruct the decay rate of the scalar, $\gf$, from the observed signal. In the next section, we introduce the template that we use for the signal, including the modified redshift due to matter domination. Subsequent sections present the analysis of the reconstruction of the spectral and thermodynamical parameters of the signal.

%~~~~~~~~~~~~~~~~~~~~~~~~~~~~~~~~~~~~~~~~~~~~~~~~~~~~~~~~~~~~~
\section{Signal template}
\label{sec:templates}
%~~~~~~~~~~~~~~~~~~~~~~~~~~~~~~~~~~~~~~~~~~~~~~~~~~~~~~~~~~~~~
Various possible mechanisms sourcing GWs during a PT have been discussed in the literature~\cite{Caprini:2015zlo,Caprini:2019egz,LISACosmologyWorkingGroup:2022jok,Caprini:2024hue}.
Following ref.~\cite{Lewicki:2022pdb}, we assume that the GW signals from very strong PTs have the same spectral shape, independently of whether they were sourced by bubble collisions or by plasma motion.~\footnote{Another possible source is associated with density fluctuations produced by the phase transition~\cite{Lewicki:2024ghw,Lewicki:2024sfw}. This effect could have a significant impact on our results as it tends to dominate the signal in the strongly supercooled regime which we also focus on. However, the exact shape of the spectrum from this source is not known yet for models with inefficient reheating and we do not take it into account in the present work.} We will use the following parameterisation~\cite{Ellis:2023oxs} for the emitted GW spectral shape,
\begin{equation}
    \label{eq:spectrum_spectral_parametrization}
  %  \Omega_{\mathrm{GW},{\rm emit}}\left(f\right)=\Omega_p
  S(f,f_p,f_*,d)=\frac{(a+b)^c }%S_*\left(f, f_*\right)}
    {\left(b\left[\frac{f}{f_p}\right]^{-\frac{a}{c}}+a\left[\frac{f}{f_p}\right]^{\frac{b}{c}}\right)^c}\left(\frac{1 + (\frac{f_*}  {f}) ^ {(a / \gamma)}}{  1 + (\frac{f_*} {f}) ^ {(d / \gamma)} } \right)^ {\gamma}\, ,
\end{equation}
where $f_p$ is the frequency of the peak and we fix the slopes following~\cite{Lewicki:2022pdb,Caprini:2024hue} as $a=b=2.4$ and $c=4$. The frequency $f_*$  corresponds to the horizon-size wavelength at the time of the phase transition~\cite{Gouttenoire:2023bqy,Ellis:2023oxs}.  For frequencies far below $f_*$ the spectrum scales universally, depending on the energy budget of the universe~\cite{Caprini:2009fx,Cai:2019cdl}. The parameter $d$ defines the slope of the spectrum in this regime, while $\gamma$ controls how quickly that asymptotic slope is reached. We will use two canonical slopes, that is, $d=3$ corresponding to standard radiation domination~\cite{Caprini:2009fx,Cai:2019cdl}~\footnote{The variation in the number of degrees of freedom modifies this simple result and also has an impact on the shape of the spectrum~\cite{Brzeminski:2022haa,Franciolini:2023wjm}. Such modifications are also not limited to GW sources associated with a PT~\cite{Witkowski:2021raz}. } and $d=1$ corresponding to a matter-domination period after the transition~\cite{Barenboim:2016mjm,Ellis:2020nnr}. The last term in the equation above parametrises our ignorance concerning the shape of the spectrum at the horizon size. In principle, a numerical simulation including the expansion of the universe would give us the exact shape. However, in the absence of such a simulation, we simply fix $\gamma =1$ for $d=3$ and $\gamma= \frac{1}{3}$ for $d=1$, respectively, which results in a quick transition not modifying the spectrum far from the breaking point at $f_*$ in each case.

\begin{figure}[ht]
    \centering
    \begin{minipage}{0.74\textwidth} 
        \centering
        \includegraphics[width=\textwidth]{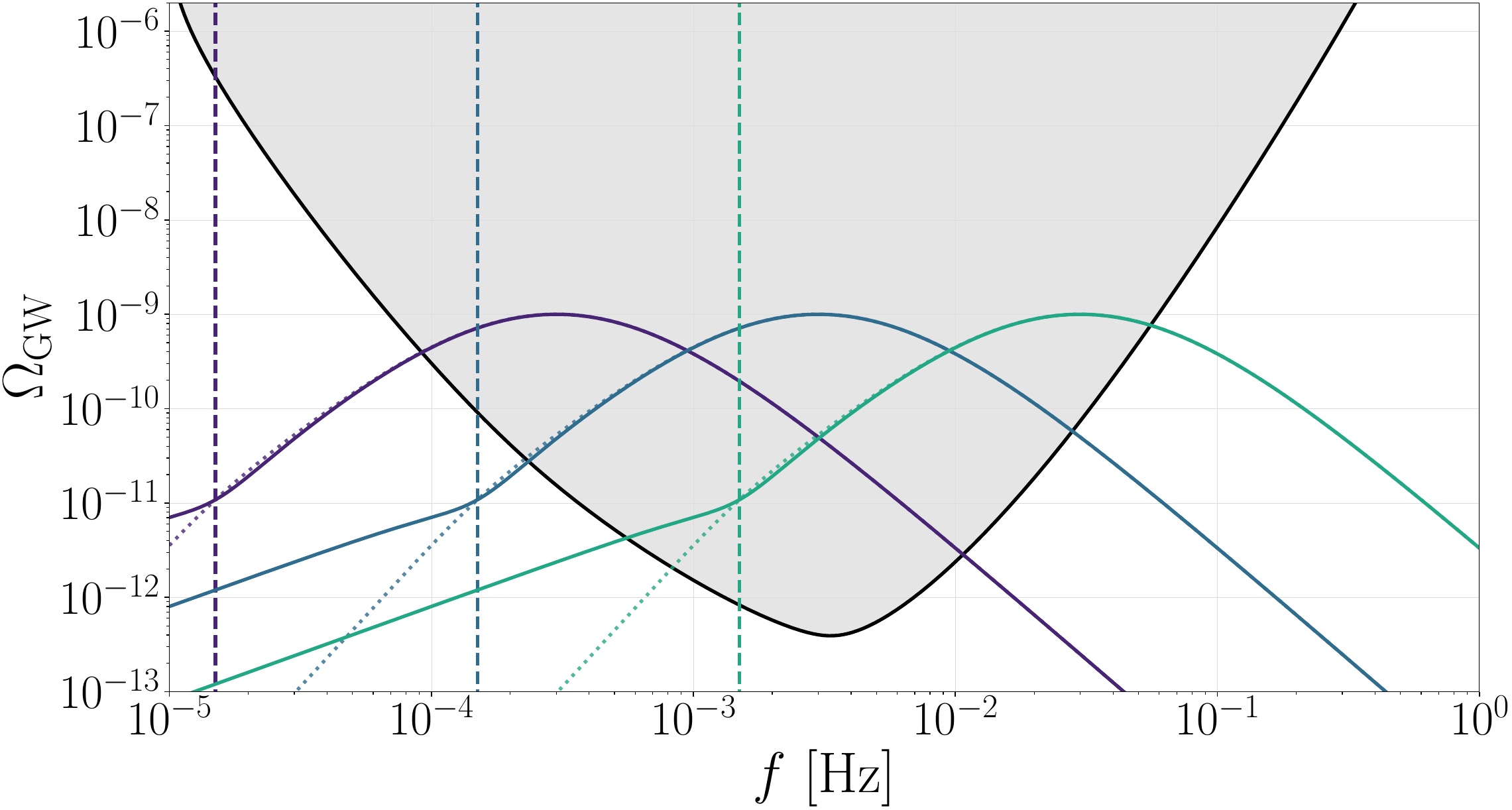}
        \vspace{2pt}
    \end{minipage}
    \begin{minipage}{0.25 \textwidth} 
        \centering
        \includegraphics[width=\textwidth]{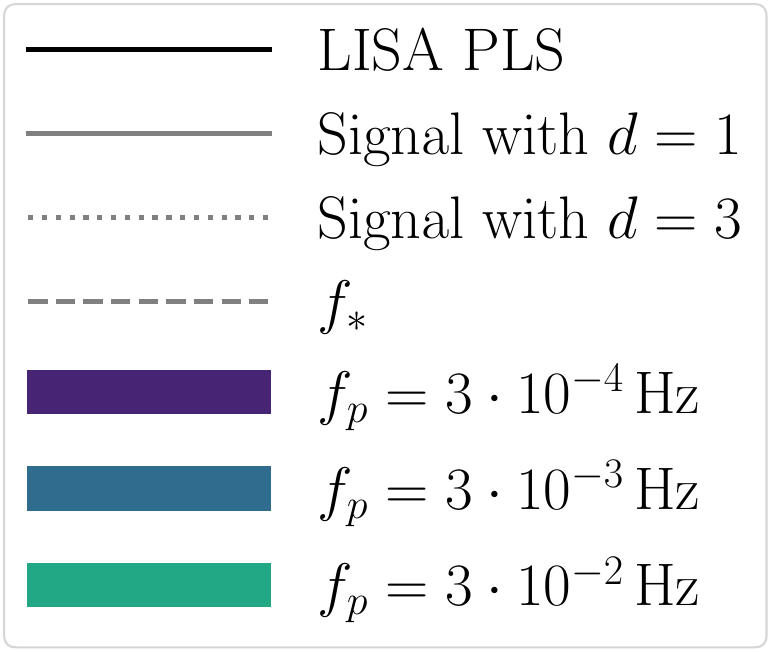}
        \vspace{2pt}
    \end{minipage} 
      \begin{minipage}{0.74\textwidth} 
        \centering
        \includegraphics[width=\textwidth]{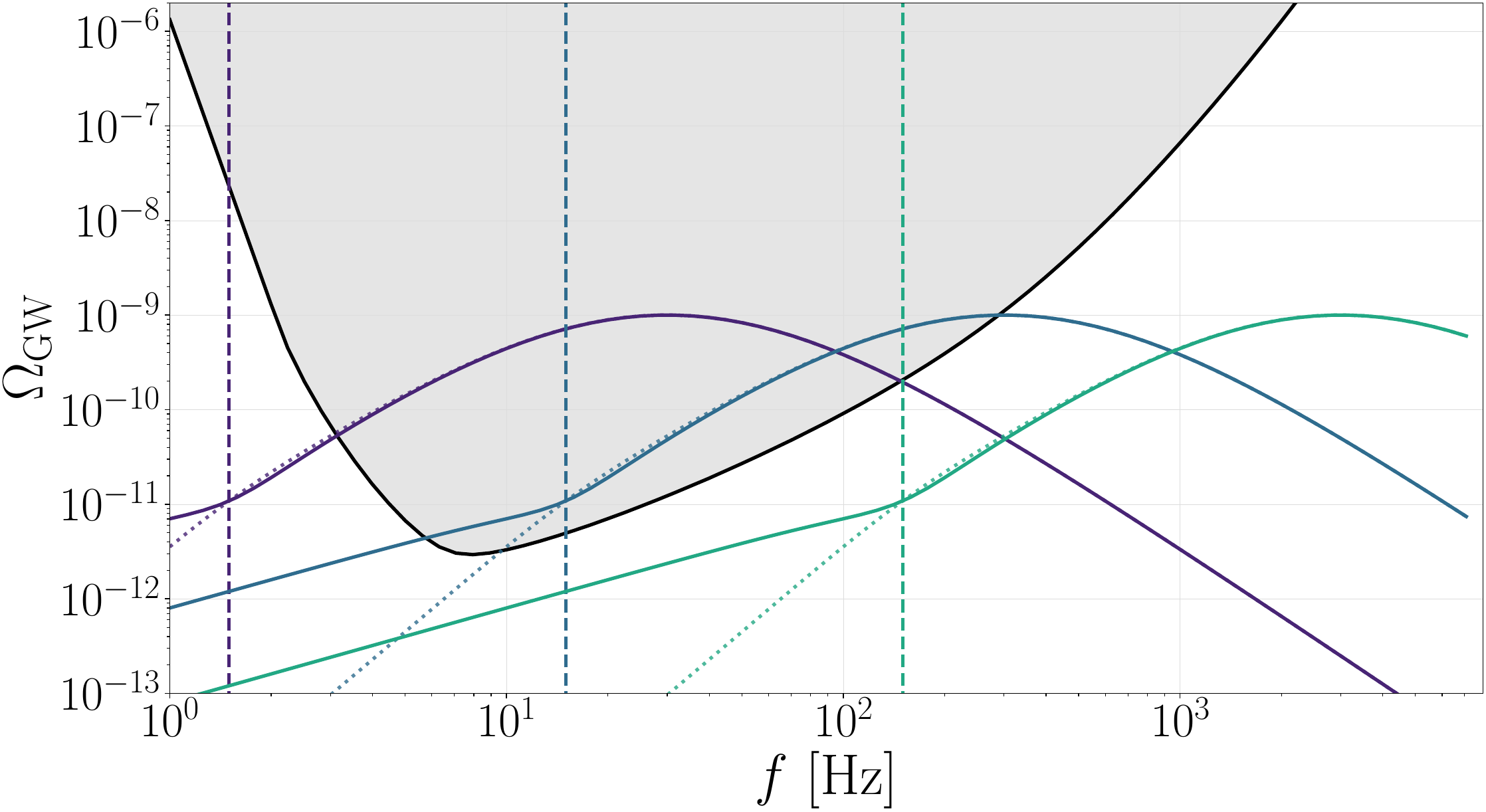}
    \end{minipage}
    \begin{minipage}{0.25 \textwidth} 
        \centering
        \includegraphics[width=\textwidth]{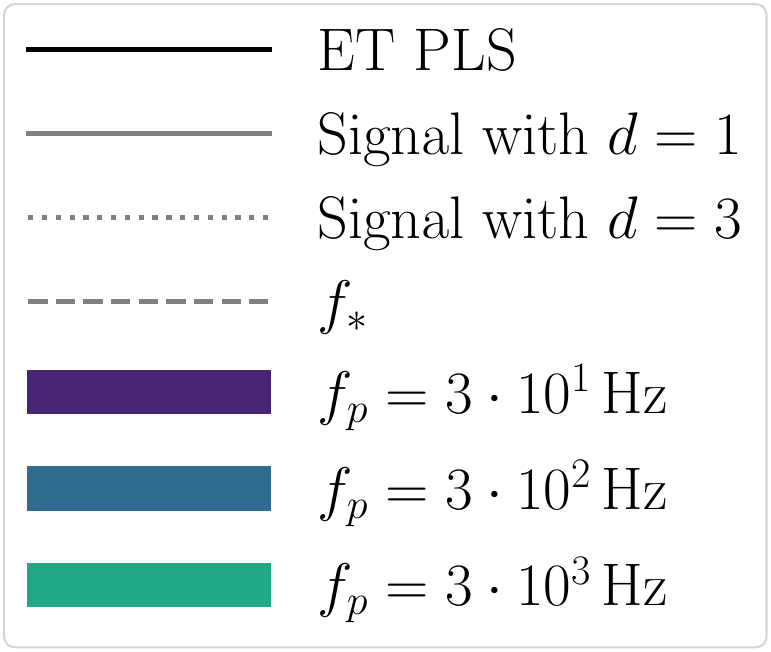}
    \end{minipage}       
\caption{Examples of spectra with fixed peak amplitude $\Omega_p=10^{-9}$, and various peak frequencies $f_p$ indicated by colour (see the legend). The breaking frequencies $f_*$ are marked by vertical dashed lines in respective colours. LISA power-law integrated sensitivity curve (PLS) is shown with a solid black line in the top panel while the bottom panel features the PLS for ET.}
    \label{fig_plot_spectra_d}
\end{figure}

In a simplified geometrical (spectral) parameterisation of the spectrum we treat the amplitude $\Omega_p$ and peak frequency $f_p$ as well as the low-frequency slope $d$ as free parameters such that the observed spectrum reads
\begin{equation}
\label{eq:Omega_geom}
    \Omega_{\rm GW}^{\rm GEOM} =  \Omega_p S(f,f_p,f_*,d)\, ,
\end{equation}
where the spectral shape is given by eq.~\eqref{eq:spectrum_spectral_parametrization}. We show example spectra of interest for LISA and ET in figure~\ref{fig_plot_spectra_d}. Solid lines show the spectra with $d=1$ corresponding to matter domination after the phase transition, while dotted lines represent $d=3$ corresponding to standard radiation domination. The peak amplitude is fixed, $\Omega_p=10^{-9}$, while different colours represent different peak frequencies. The breaking frequencies, $f_*$ are marked by vertical dashed lines.~\footnote{We explain how to evaluate $f_*$ below.}

In order to compute the spectrum for a particular particle physics model we need to relate the spectrum to the parameters defining the transition which we discussed in section~\ref{sec:PT}. This gives the emitted spectrum
\begin{equation}
\label{eq:Omega_th}
    \Omega_{\rm GW, \rm \,emit}^{\rm TH} =  5.1\cdot 10^{-2} \left(\frac{\beta}{H_*}\right)^{-2} S(f,f_p,f_*,3), \quad f_p=0.7 f_*\frac{\beta}{H_*},
\end{equation}
where the numerical factors are determined through numerical simulations~\cite{Lewicki:2022pdb}. This gives us the spectrum at the time of the production and in order to compute the spectrum today we need to properly include the redshift. This can be done using~\cite{Ellis:2020nnr}
\begin{equation}
\label{eq:amplitude_redshift}
   \Omega_\text{GW}^{\rm TH}= \Omega^{\rm TH}_{\text{GW},\,\text{emit}} \times \begin{cases}
 \left(\frac{a_{*}}{a_0}\right)^4\left(\frac{H_{*}}{H_0}\right)^2,\quad {\rm for} \quad f \geq f_*\\
      \left(\frac{a_f}{a_0}\right)^4\left(\frac{H_f}{H_0}\right)^2,\quad {\rm for} \quad f<f_*
   \end{cases}
\end{equation}
where the lower case describes the modes that are super-horizon at the time of the transition, which begin to redshift only once they enter the horizon at $a=a_f$, $H=H_f$. Above we also use the standard relation between scales and frequencies for the horizon size at the time of the transition and at the moment when the mode of a generic frequency $f$ re-enters the horizon
\begin{equation}
    f_*=\frac{a_* H_*}{2\pi}\,, \quad f=\frac{a_f H_f}{2\pi} \text{.}
\end{equation}
In the standard radiation domination case eq.~\eqref{eq:amplitude_redshift} simply gives $\Omega_{\rm GW}^{\rm GEOM}=1.67 \cdot 10^{-5}\, \Omega_{\rm GW, \, emit}^{\rm TH} $.

\begin{figure}[t]
    \centering
    \begin{minipage}{0.74\textwidth} % Width for the first figure
        \centering
        \includegraphics[width=\textwidth]{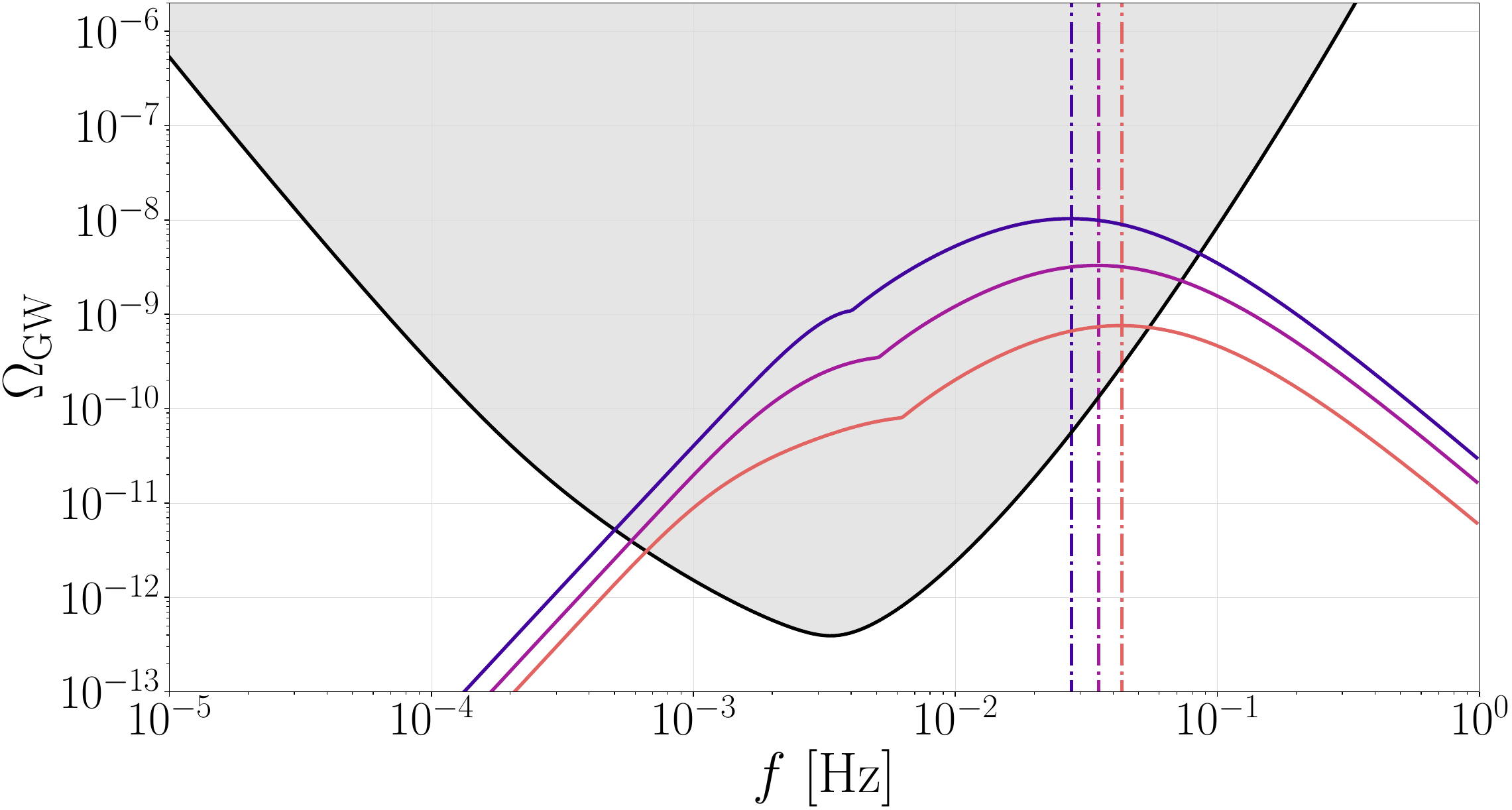}
        \vspace{2pt}
    \end{minipage}
    \begin{minipage}{0.25\textwidth} % Width for the second figure
        \centering
        \includegraphics[width=\textwidth]{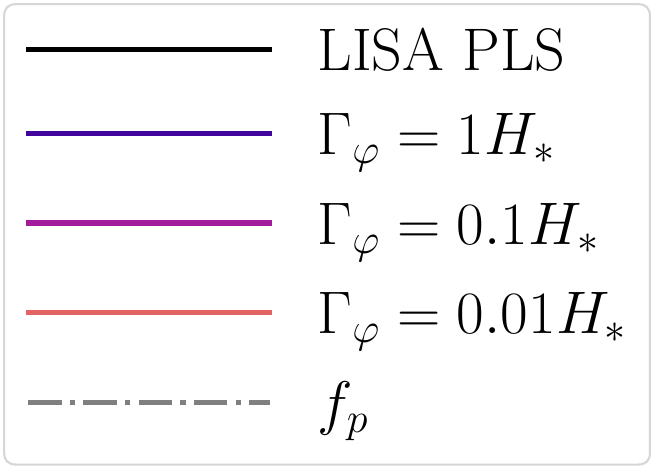}
        \vspace{2pt}
    \end{minipage} 
        \begin{minipage}{0.74\textwidth} % Width for the first figure
        \centering
        \includegraphics[width=\textwidth]{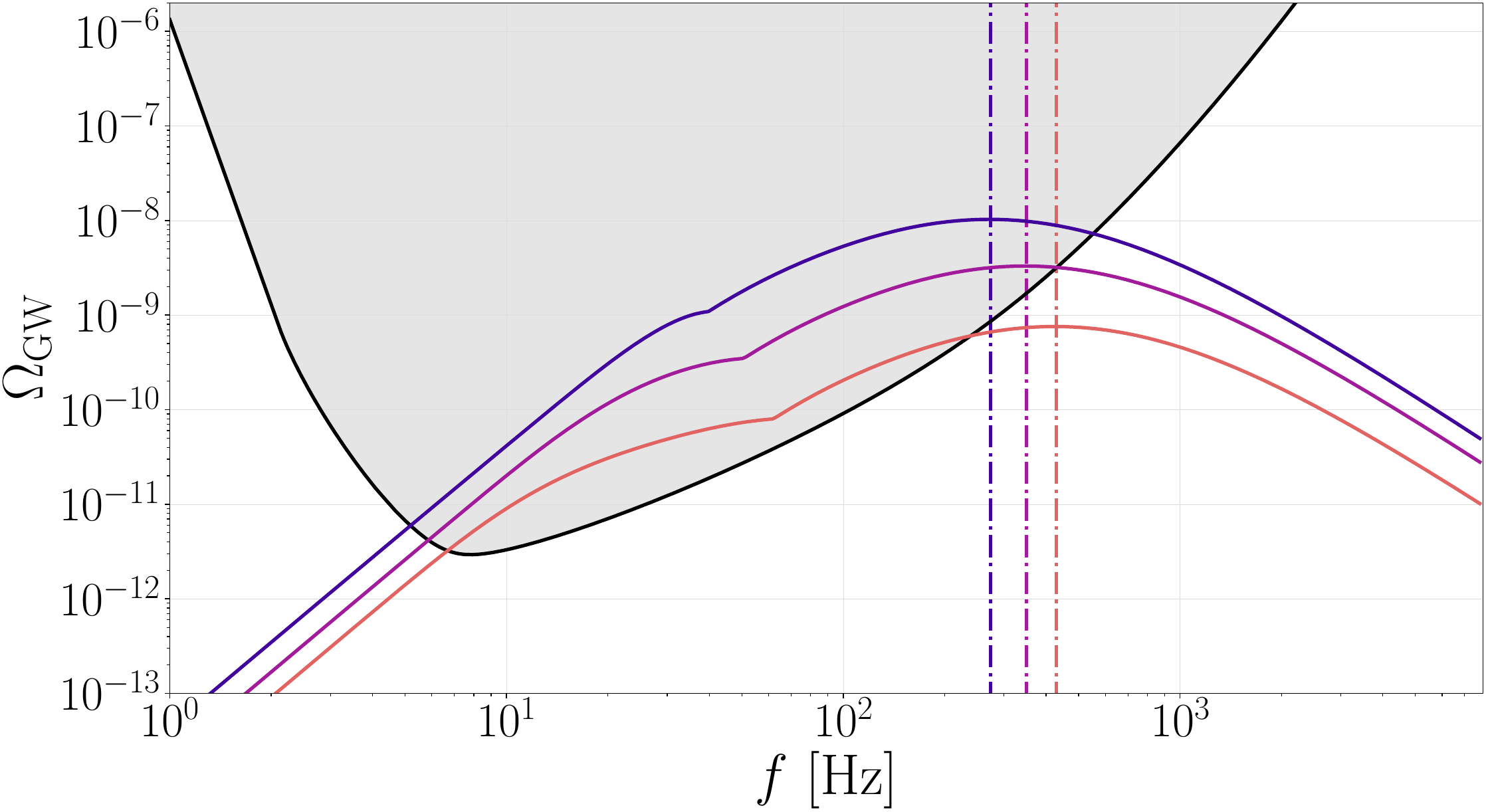}
    \end{minipage}
    \begin{minipage}{0.25\textwidth} % Width for the second figure
        \centering
        \includegraphics[width=\textwidth]{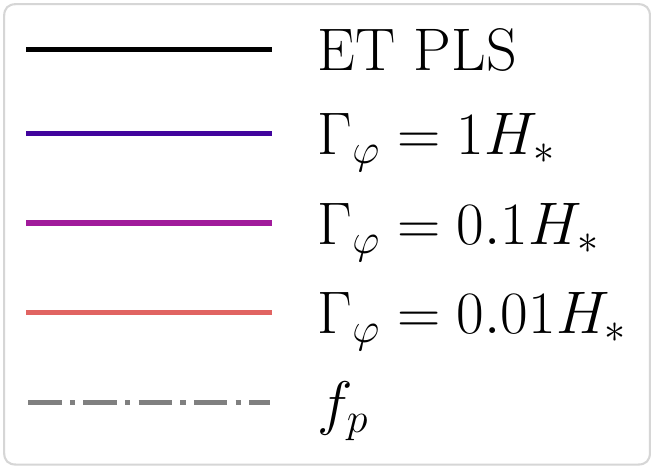}
    \end{minipage}        
 \caption{The solid lines show example spectra for the three benchmark values for $\gf=H_*,\  0.1H_*,\  0.01H_*$ as indicated by the colours. Peak frequencies are marked with dash-dotted vertical lines in respective colours while the solid black lines correspond to the PLS sensitivity curves. In the top panel, the parameters are fixed to $T_{\text{max}}=10^5\text{ GeV}$ and $\beta/H_*=10$ with the PLS curve corresponding to LISA while in the bottom panel $T_{\text{max}}=10^9\text{ GeV}$ and $\beta/H_*=10$ and the PLS curve corresponds to ET. }
    \label{fig_plot_spectra_gmma}
\end{figure}

Unlike in the simple radiation-dominated case, in order to include the redshift we need to solve the Boltzmann equations~\eqref{eq:Boltzman} to find the evolution of the scale factor included in the equations above. In this way, we account for the dilution of the signal due to a period of matter domination and the influence of the scalar decay rate  $\Gamma_\varphi$ on the spectrum. To demonstrate this, we will focus on three benchmark values of $\gf$ which we already discussed in section~\ref{sec:PT} and for which the evolution is depicted in figure~\ref{fig:solutions-Boltzmann}. Figure~\ref{fig_plot_spectra_gmma} shows the resulting spectra with the colour indicating the value of $\gf$, while all other parameters are fixed. First, we see the plateau at low frequencies now clearly depending on the length of the matter domination period. The smaller the decay rate the longer it takes before the field decays into radiation. Thus, the corresponding plateau at frequencies below the horizon size is longer for smaller values of $\gf$. Now we carefully treat the reheating after the PT and model the redshift in agreement with the actual temperature evolution of the universe after the PT, as described in section~\ref{sec:PT}. Therefore, we can see the impact of modified redshift due to the matter domination on the part of the spectrum within the horizon at the time of transition. The modified redshift results in diminished peak amplitude and somewhat increased peak frequency for a longer period of matter domination.

Relating the geometrical and the thermodynamical parameterizations, we can evaluate the first break, $f_*$ in order to only look at physically relevant parameters.
 Using the above simplification, with eqs.~\eqref{eq:Omega_geom} and \eqref{eq:Omega_th} substituted, and evaluated at peak frequency, such that $S\approx 1$, we can eliminate $\beta/H_*$ to find $f_*$,
\begin{equation}
\label{f_*_spectral}
  f_* \approx \frac{ 1 }{0.7 \sqrt{ 5.1\cdot 10^{-2}\cdot 1.67\cdot 10^{-5}}} \sqrt{\Omega_p} f_p \approx  1.56\cdot 10^{3} \sqrt{\Omega_p} f_p\, .
\end{equation}
Due to the simplification, this expression holds for the pure radiation case of $d=3$ and can be used in eq.~\eqref{eq:Omega_geom}. While reconstructing signals from the geometric parameters we will also use it for $d=1$, implicitly assuming that the matter domination period lasts for a very short time such that the modification of the peak due to the redshift can be neglected.

%~~~~~~~~~~~~~~~~~~~~~~~~~~~~~~~~~~~~~~~~~~~~~~~~~~~~~~~~~~~~~
\section{Detector noise}\label{sec:noise}
%~~~~~~~~~~~~~~~~~~~~~~~~~~~~~~~~~~~~~~~~~~~~~~~~~~~~~~~~~~~~~
In this section, we will review the noise models for the experiments we focus on. We start with LISA using a noise model based on refs.~\cite{Babak:2021mhe} and~\cite{Karnesis:2021tsh}. The main sources of noise are acceleration and path-length uncertainty, defined by the characteristic amplitudes
\begin{equation}
\sqrt{(\delta a)^2}=3 \cdot 10^{-15} \mathrm{~m} / \mathrm{s}^2, \quad \sqrt{(\delta x)^2}=1.5 \cdot 10^{-11} \mathrm{~m}  \text{.}
\end{equation}
They combine to the effective noise power spectral density~\cite{Babak:2021mhe}
\begin{equation}
\label{eq:S_eff}
   S_{\text {noise }}^{\text {eff,LISA }}(f)  =\frac{10}{3}\left(\frac{S_I(f)}{(2 \pi f)^4}+S_{I I}\right) R(f) +S_{\text{Gal}}(f)\text{,}
\end{equation}
where $S_I$ and $S_{II}$ encode the two sources of noise and are given by
\begin{equation}
\begin{aligned}
\label{SS}
S_I & =4\left(\sqrt{(\delta a)^2} / L\right)^2\left(1+\left(f_1 / f\right)^2\right) \mathrm{Hz}^{-1}=5.76 \cdot 10^{-48}\left(1+\left(f_1 / f\right)^2\right) \mathrm{s}^{-4} \mathrm{~Hz}^{-1} \text{,}\\
S_{I I} & =\left(\sqrt{(\delta x)^2} / L\right)^2 \mathrm{~Hz}^{-1}=3.6 \cdot 10^{-41} \mathrm{~Hz}^{-1} \text{,}
\end{aligned}
\end{equation}
while the inverse response function reads
\begin{equation}
\label{eq:R(f)}
    R(f)=\left(1+\left(\frac{f}{f_2}\right)^2\right)\text{.}
\end{equation}
We use the length of the LISA arm $L=2.5\cdot10^9\, \mathrm{m}$ and characteristic frequencies $f_{1}=0.4\, \mathrm{mHz}$ and $f_{2} = 25\ \mathrm{mHz}$ \cite{Babak:2021mhe}. The last term $S_{\text{Gal}}$ in eq.~\eqref{eq:S_eff} refers to galactic confusion noise produced by binary white dwarfs which can be approximated by the analytic fit~\cite{Karnesis:2021tsh},
\begin{equation}
S_{\mathrm{Gal}}(f)=A f^{-7 / 3} e^{-\left(\frac{f}{f_3}\right)^\alpha} \frac{1}{2}\left[1.0+\tanh \left(-\frac{f-f_k}{f_4}\right)\right]\text{,}
\end{equation}
where $A=1.14 \cdot 10^{-44},\, \alpha=1.8, f_4=0.31\,\mathrm{mHz}$ and $f_3, f_k$ depend on the observation time $T_{\mathrm{obs}}$, via
\begin{equation}
\log _{10}\left(f_3\right)=a_3 \log _{10}\left(T_{\mathrm{obs}}\right)+b_3, \quad \log _{10}\left(f_k\right)=a_k \log _{10}\left(T_{\mathrm{obs}}\right)+b_k \text{.}
\end{equation}
Here the unit of $T_{\mathrm{obs}}$ is years and $a_3=-0.25$, $b_3=-2.7$ and $a_k=-0.27$, $b_k=-2.47$.

Next, let us turn to the sensitivity of the Einstein Telescope. The final design of the experiment is still debated~\cite{Branchesi:2023mws}.
For simplicity, we use the interpolated function for the power spectral density $D_{\text {noise }}^{\mathrm{ET}}$, for the D design \cite{ET_Link, Hild:2010id}.
The effective noise power spectral density can be written as~\cite{Schmitz:2020syl}:
 \begin{equation}
    \label{eq:S_eff_ET}
    S_{\text {noise }}^{\text {eff,ET }}(f)=\frac{1}{2}\frac{1}{\sqrt{3}}\left|\frac{D_{\text {noise }}^{\mathrm{ET}}(f)}{\Gamma_{\mathrm{ET}}(f)}\right| \text{,}
\end{equation}
where ${\Gamma_{\mathrm{ET}}(f)}$ given in \cite{Regimbau:2012ir}  is the overlap reduction function and we include an additional factor of one half, to follow our convention for equation \eqref{eq:Simga_Omega}~\footnote{See eqs. (A.55) and (A.24) in~\cite{Schmitz:2020syl} for details.}.

%~~~~~~~~~~~~~~~~~~~~~~~~~~~~~~~~~~~~~~~~~~~~~~~~~~~~~~~~~~~~~
\section{Fisher matrix reconstruction}\label{sec:fisher}
%~~~~~~~~~~~~~~~~~~~~~~~~~~~~~~~~~~~~~~~~~~~~~~~~~~~~~~~~~~~~~

For a given spectrum $\Omega_{\rm GW}$ characterised by parameters $\theta_{\alpha}$, the Fisher information matrix allows one to evaluate the accuracy of reconstruction of $\theta_{\alpha}$ values from the observed signal, under the assumption of Gaussian likelihood function \cite{Fisher:1922saa, Vallisneri:2007ev}. 
An element of the Fisher information matrix is given by~\cite{Caprini:2024hue,Gowling:2021gcy,Smith:2019,Gowling:2023ocq,Blanco-Pillado:2024aca},
\begin{equation}
\label{Fisher_eq}
    F_{\alpha\beta} =  T_{\text{obs}} \bigintssss_{f_{\text{min}}}^{f_{\text{max}}} \dd f \frac{\pdv{\Omega_{\rm GW}}{\theta_{\alpha}}\pdv{\Omega_{\rm GW}}{\theta_{\beta}}}{\Big(\Omega_{\text{noise}}+%\frac{3H_0^2}{4\pi^2 f^3}\Omega_{\rm GW}(f)S_{\text{eff}}
    \Omega_{\rm GW}\Big)^2} \\, 
\end{equation}
where we converted the effective noise $S_{\text{eff}}$ from eq.~\eqref{eq:S_eff} into fractional energy density parameter
\begin{equation}
\label{eq:Simga_Omega}
   \Omega_{\text{noise}}=\frac{4 \pi^2 f^3}{3 H_0^2} S_{\text{eff}}\, .
\end{equation}
We note that in the expression above we assumed that the $A$ and $E$ channels give equal contributions. We also assume $T_{\text{obs}}=4$ years, for both LISA and ET.

The inverse of the Fisher matrix gives the squared covariance matrix
\begin{equation}
    \sigma^2=F^{-1}\, .
\end{equation}
The diagonal elements, treated separately, give projected squared uncertainties of the reconstruction of a given parameter when we marginalise over other parameters. The off-diagonal elements represent covariance between different parameters. One can diagonalise the covariance matrix to obtain new variables whose uncertainties are uncorrelated.
Let us define the (relative) normalized uncertainties using the following relations
\begin{equation}
\label{eq:deltas}
\delta_{xy}=
\begin{cases}
\frac{\sigma_{xx}}{x} \text{ for }x=y\,, \\
\frac{\sqrt{\left|\sigma_{xy}^2\right|}}{\sqrt{xy}} \text{ for }x\neq y\, ,
\end{cases}
\end{equation}
where we write every element of covariance matrix $\sigma^2$ matrix as $\sigma_{xy}^2$. Note that non-diagonal elements can be negative, so we introduced absolute value in \eqref{eq:deltas}.
%Normalized uncertainties are identical to relative uncertainties,
Mathematically, this definition can be understood as applying the natural logarithm to the parameters of the covariance matrix, i.e.~
$\theta_i \to \ln \theta_i$, then $\sigma^2_{\ln \theta_i, \ln \theta_j}=\frac{\sigma^2_{\theta_i\theta_j}}{\theta_i \theta_j}$, see section 4.1 of ref.~\cite{Gowling:2021gcy} or section 3.4 of ref. \cite{Caprini:2024hue}.

Let us also define the signal-to-noise ratio $\text{SNR}$~\cite{Caprini:2015zlo} 
\begin{equation}
\label{eqn:SNRdef}
\text{SNR}=\Bigg[T_{\mathrm{obs}}\int%_{f_{\text{min}}}^{f_{\text{max}}}
\dd f\frac{\Omega_{\text{GW}}^2}
{\Omega^2_{\text{noise}}}\Bigg]^{\frac{1}{2}}\, ,
\end{equation}
where the integral runs over the sensitivity window of the experiment. The usual criterion of observability, $\text{SNR}=10$, will give us a reference in the analysis of the reconstruction of parameters.

Let us note that the Fisher formalism can break down for signals that have low SNR, or signals that weakly depend on certain parameters, see ref.~\cite{Vallisneri:2007ev}. This could be improved by performing a full Bayesian analysis \cite{Toubiana:2020vtf, Caprini:2024hue}, which, however, is out of the scope of this paper. We also note that for large SNR the calculation of SNR becomes unreliable and gives only an upper limit, due to intrinsic noise, see ref.~\cite{Allen:1997ad}.

%~~~~~~~~~~~~~~~~~~~~~~~~~~~~~~~~~~~~~~~~~~~~~~~~~~~~~~~~~~~~~
%%%%%%%%%%%%%%%%%%%%%%%%%%%%%%%%%%%%%%%%%%%%%%%%%%%%%%%%%%%%%%%%%%%%%
%%%%%%%%%%%%%%%%%%%%%%%%%%%%%%%%%%%%%%%%%%%%%%%%%%%%%%%%%%%%%%%%%%%%%
\section{Reconstruction of spectral parameters from the signal}\label{sec:reconstruction-spectral}

\begin{figure}[t]
    \centering
    LISA \\[2pt]
    \includegraphics[width=0.85\linewidth]{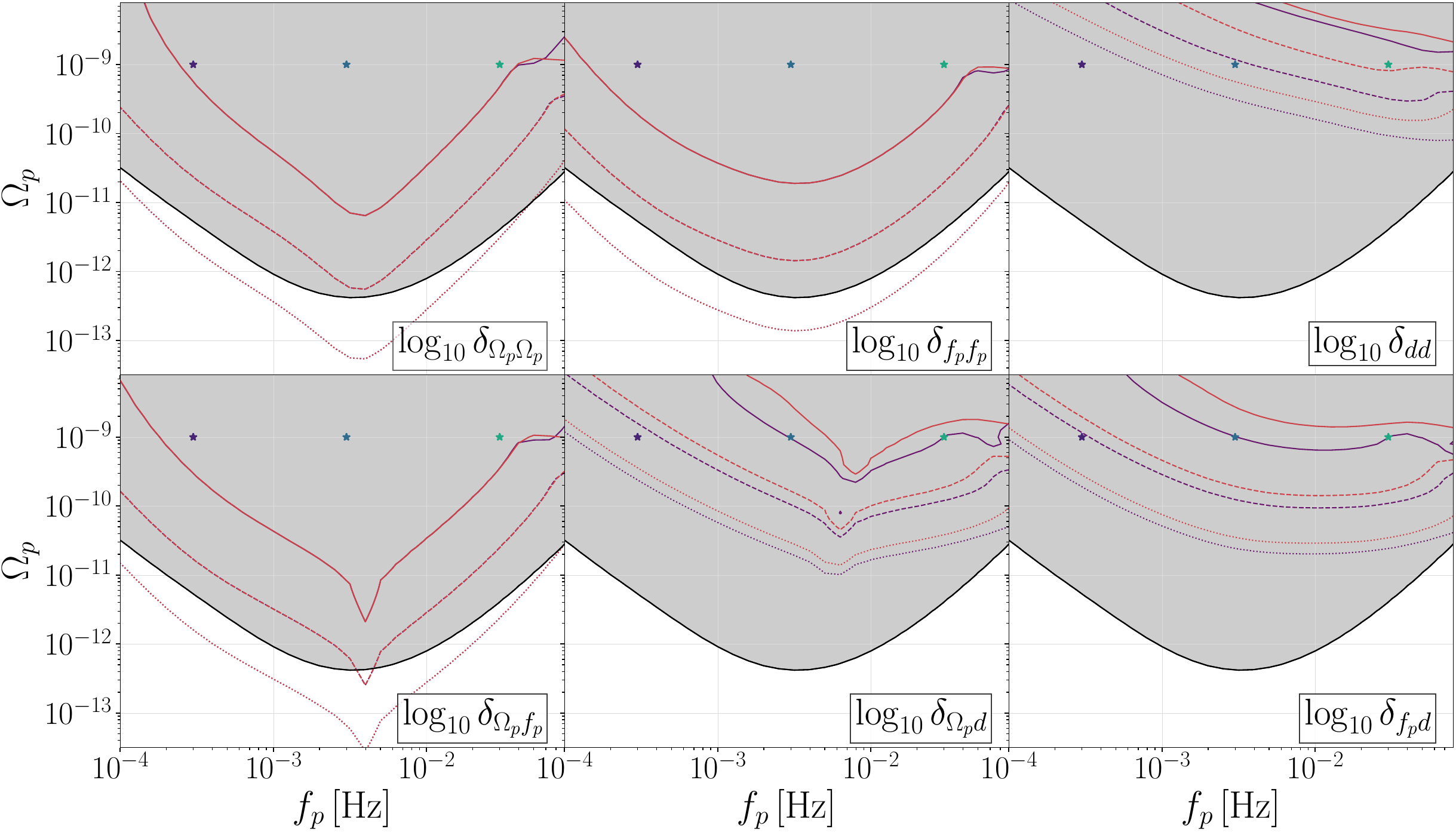}\\[2pt]
    ET \\[2pt]
   \includegraphics[width=0.85\linewidth]{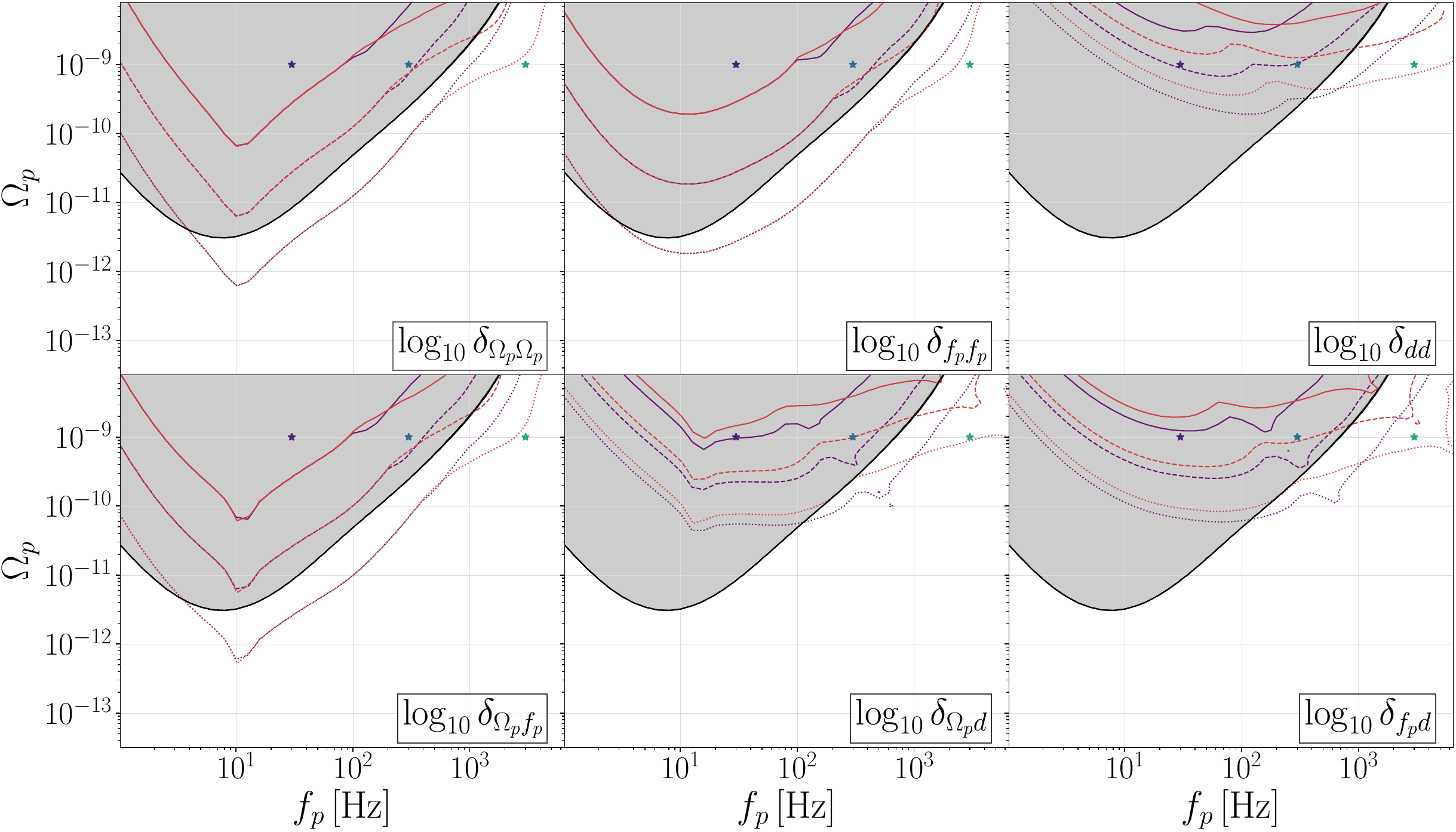}\\[2pt]
    \includegraphics[width=0.65\textwidth]{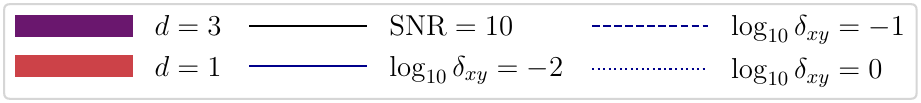}
    \caption{
Relative uncertainties in the reconstruction of geometrical parameters with LISA (top panel) and ET (bottom panel), for the low-frequency slope $d=3$ and $d=1$ (see eq.~\eqref{eq:spectrum_spectral_parametrization}). Solid, dashed and dotted blue lines correspond to the normalised uncertainties of $1\%$, $10\%$ and $100\%$ respectively. The black solid line corresponds to ${\rm SNR}=10$ for the signal with d=3 for comparison.  The stars in the top and bottom panels represent parameters of the spectra from the top and bottom panels of figure~\ref{fig_plot_spectra_d}.
}
    \label{fig_d3_uncertainty}
\end{figure}

We will begin with the reconstruction of geometrical parameters that is the peak amplitude $\Omega_{p}$, frequency $f_{p}$ and low-frequency slope parameter $d$ as introduced in eq.~\eqref{eq:spectrum_spectral_parametrization}.
Figure~\ref{fig_d3_uncertainty} shows normalised uncertainties of the reconstruction of these parameters as functions of the amplitude and peak frequency for fixed low-frequency slope $d$. Solid, dashed and dotted lines show three contours corresponding to the uncertainty of $1\%$, $10\%$ and $100\%$ of the value, respectively.
Purple contours correspond to $d=3$ while the red ones to $d=1$. In the top panel, we show the reconstruction results for LISA while in the lower panel for ET. The black line shows the simplified detection criterion for our spectra using $\text{SNR}=10$ for $d=3$. The stars in both panels correspond to the spectra shown in the two panels of figure~\ref{fig_plot_spectra_d} with the colours indicating the parameters defining the given spectrum. For the parameters defining the amplitude and position of the peak we find that, as expected, the $\text{SNR}=10$ line falls between $10\%$ and $100\%$ uncertainty on parameters. The same is not true for the low-frequency slope parameter $d$ which requires a very large amplitude of the signal and peak frequency somewhat above the peak sensitivity of the experiment. This is very simple to understand by looking at the example spectra in figure~\ref{fig_plot_spectra_d}. The modification below the frequency corresponding to the horizon size appears at frequencies significantly below the peak. The amplitude of the signal at that frequency is also significantly lower and only if the peak is well within the sensitivity the modification can be observed. Nonetheless, we find a viable region in the parameter space where the evolution history of the universe can be probed. In particular, in the region where $\delta_{dd}<\mathcal{O}(10-100\%)$, the scenarios with and without early matter domination can be distinguished.

The analysis presented here is subject to certain simplifications, mentioned before. The $d=3$ case corresponds to the standard scenario of radiation domination while $d=1$ corresponds to a period of matter domination after the transition. In this example, we keep the redshift as in the standard radiation case also for $d=1$ which leads again to eq.~\eqref{f_*_spectral}. A prolonged period of matter domination leading to an extended plateau with $\Omega\propto f$ for frequencies below the horizon size would lead to extra dilution according to eq.~\eqref{eq:amplitude_redshift}. Thus our example corresponds to a very brief matter domination, although the length of the plateau in the simplified spectrum does not reflect that. Therefore, the results in this plot show the most optimistic range in which the matter-domination-induced modification can be reconstructed. The analysis of thermodynamical parameters presented in the following section is free from the simplifications discussed above.

%~~~~~~~~~~~~~~~~~~~~~~~~~~~~~~~~~~~~~~~~~~~~~~~~~~~~~~~~~~~~~
\section{Reconstruction of thermodynamical parameters from the signal}\label{sec:reconstruction:thermo}
%~~~~~~~~~~~~~~~~~~~~~~~~~~~~~~~~~~~~~~~~~~~~~~~~~~~~~~~~~~~~~
For the thermodynamical parameters that fully determine the signal we choose the maximal temperature attained during the reheating process, $T_{\text{max}}$, the inverse time scale of the transition, $\beta/H_*$, and the decay rate of the scalar field, $\gf$. Their relations with the spectral parameters are discussed in section~\ref{sec:templates}. The simplified templates discussed in the previous section proved enough to verify standard expansion after the PT, for strong enough signals. However, the thermodynamical parameters have a more direct connection to the underlying model and give us a chance to understand the nature of the PT that gave origin to the observed GW signal. Reconstruction of $T_{\text{max}}$ and $\beta/H_*$ is not enough to deduce the underlying model. The spectra are degenerate with respect to those and many models lead to similar thermodynamical parameters. However, $\gf$ carries direct information about the underlying fundamental physics. The weaker the coupling between the scalar that undergoes the transition to the SM, the lower the value of $\gf$. An extended period of early matter domination, resulting in a characteristic feature in the GW spectrum, would be a clear sign of a transition in a sector weakly coupled to the SM.

\begin{figure}[t]
    \centering
    LISA\\[2pt]
        \includegraphics[width=.87\linewidth]{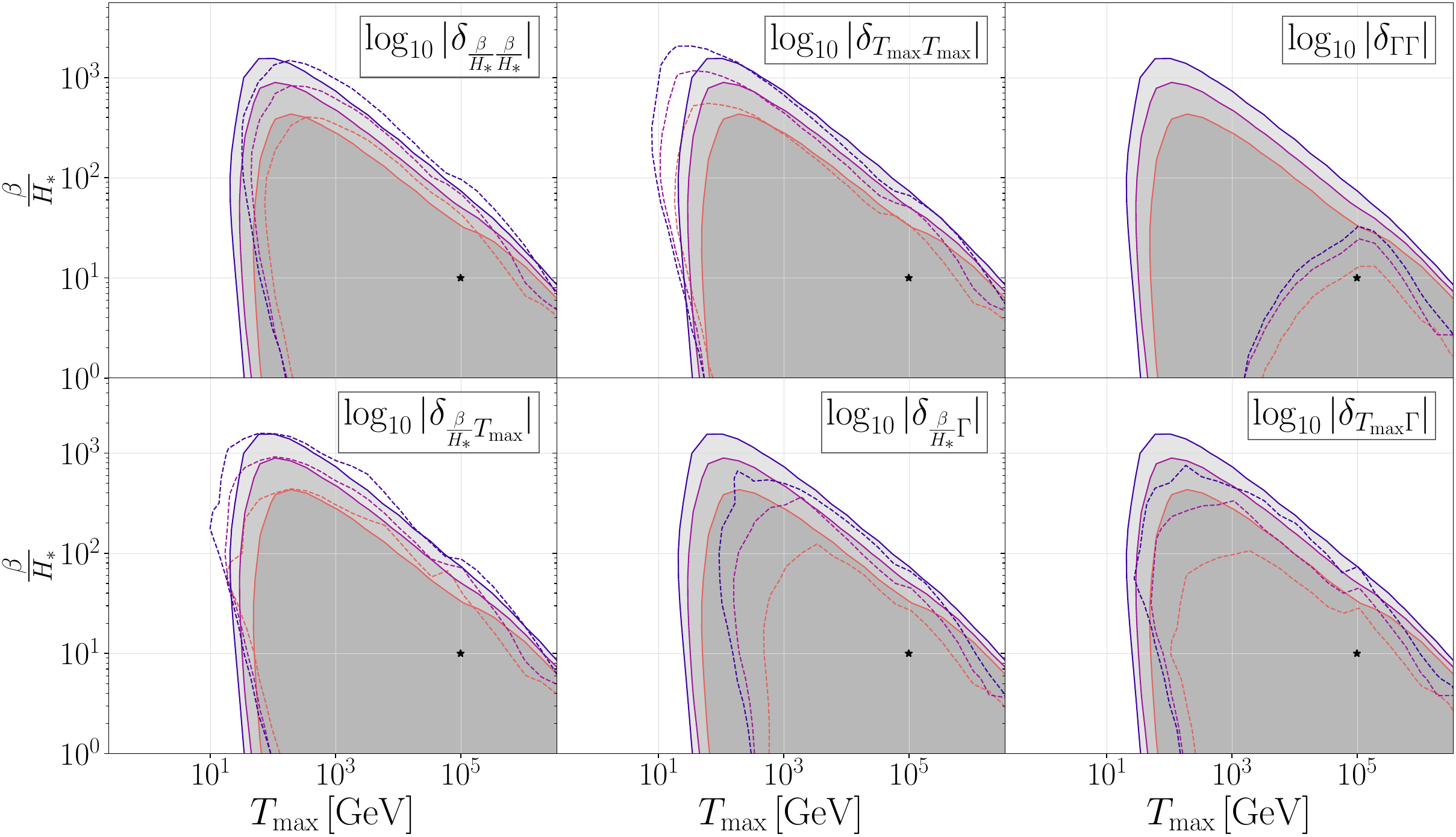}\\[2pt]
        ET\\[2pt]
                \includegraphics[width=.87\linewidth]{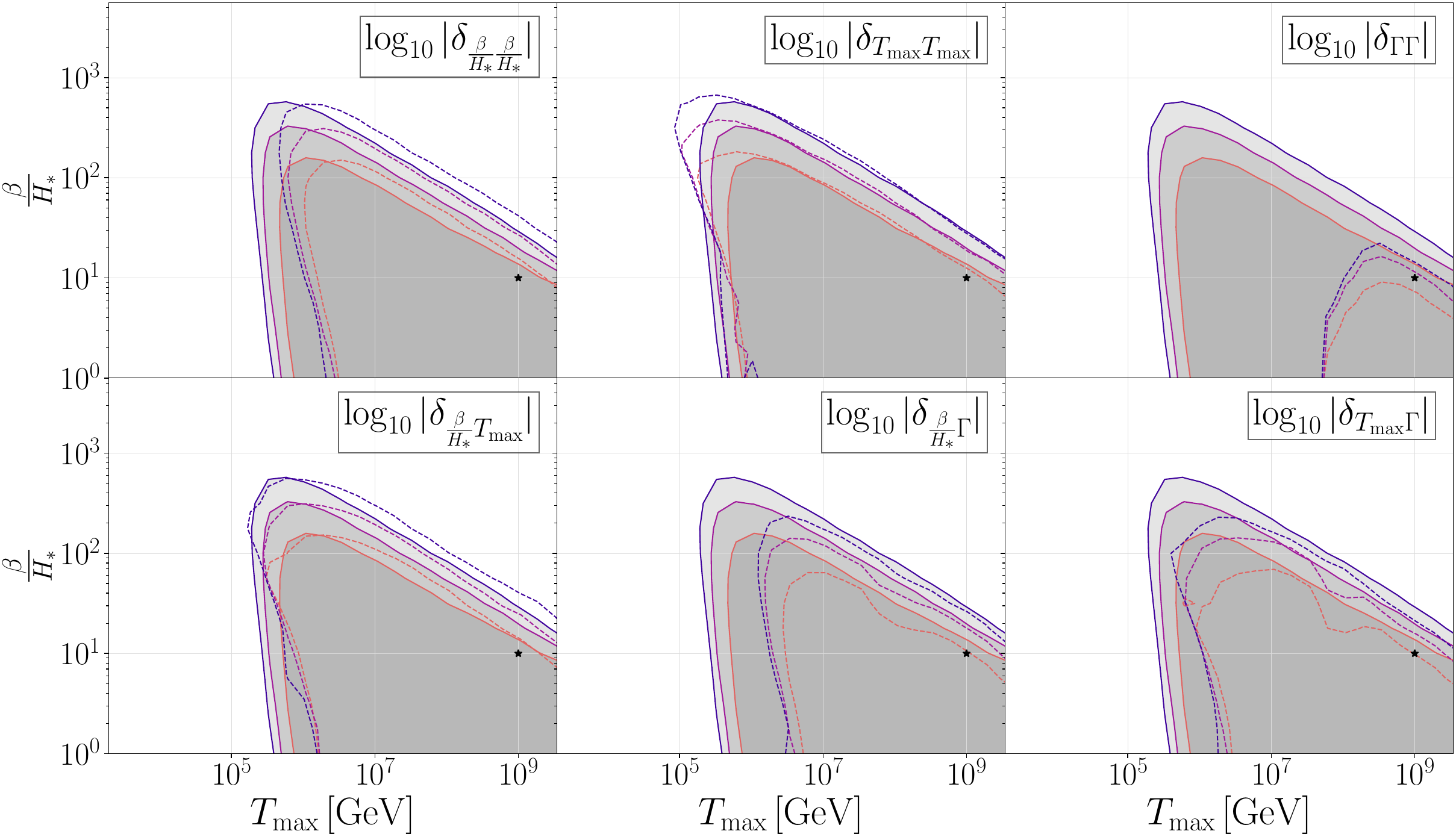}\\[2pt]
    \includegraphics[width=0.5\textwidth]{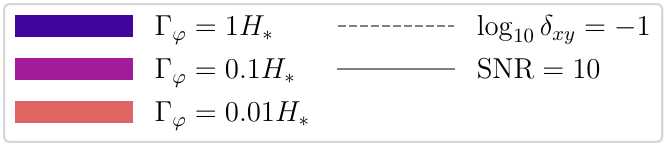}
    \caption{Reconstruction of thermodynamical parameters using LISA (top panel) and ET (bottom panel). The dashed contours correspond to a 10$\%$ relative uncertainty in reconstruction while the solid contours correspond to ${\rm SNR}=10$. The colour of the contours indicates one of the three benchmark values for $\gf/H_*=1,\  0.1, {\rm and } \  0.01$. The stars represent parameters of the spectra from the corresponding panels in  figure~\ref{fig_plot_spectra_gmma}.}
    \label{fig_gamma_uncertainty}
\end{figure}

Let us proceed with the analysis of the parameter space. Figure~\ref{fig_gamma_uncertainty} shows the relative uncertainties (see eq.~\ref{eq:deltas}) of the inverse time scale of the transition $\beta/H_*$, the maximal temperature reached $T_{\rm max}$ and the decay rate $\gf$. We again fix the value $\gf$ to our three benchmarks shown with different colours. The solid lines show the standard detection criterion $\rm{SNR}=10$ while the dashed lines show the $10\%$ uncertainties of the parameters indicated in each panel. The top panel shows the reconstruction with LISA while the bottom one with ET. Finally, the grey stars correspond to the spectra depicted in the respective panels in figure~\ref{fig_plot_spectra_gmma} (they differ only by the value of $\gf/H_*$, therefore they correspond to a single point in the space of $\beta/H_*$ and $T_{\rm max}$). As expected we see the regions where $\beta/H_*$ and $T_{\rm max}$ can be determined are similar and strongly correlated. This is due to the degeneracy as the time scale determines both the amplitude and peak frequency of the signal.
Note that the line of 10\% relative uncertainty for $T_{\rm{max}}$ is placed slightly to the left, while for $\beta/H_*$  to the right compared to $\rm{SNR} = 10$.

Most importantly we again see that the region in which the decay rate can be determined is much smaller as an appropriately strong signal is required. This is analogous to the determination of the slope $d$ in the simplified spectra from the previous section. Now we can see why ideally the matter domination period should not last too long as that diminishes the amplitude of the spectrum due to modified redshift and decreases the experimental prospects accordingly. Nonetheless, in a certain region of the parameter space $\gf$, which is a property of the underlying fundamental physics model, can be probed via GW. It may be a unique way of probing weak scalar couplings.

One may ask whether the ranges of $\beta/H_*$ and $T_{\rm{max}}$ are realistic and can be attained in physically motivated models. Popular extensions of the SM featuring supercooling are models with classical scale symmetry. Correct pattern of symmetry breaking can be easily achieved in models where classically conformal SM is extended with a new gauge group, e.g.~U(1)~\cite{Jinno:2016knw,Marzola:2017jzl,Marzo:2018nov,Baldes:2018emh,Fujikura:2019oyi,Lewicki:2021xku,Schmitt:2024pby,Lewicki:2024sfw} or SU(2)~\cite{Hambye:2013, Carone:2013, Khoze:2014, Pelaggi:2014wba, Karam:2015, Plascencia:2016, Chataignier:2018RSB, Hambye:2018qjv, Baldes:2018emh, Prokopec:2018, Marfatia:2020, Kierkla:2022odc, Kierkla:2023von}. In the model with U(1)$_{\rm B-L}$ symmetry low values of $\beta/ H_*$ are naturally obtained for reheating temperatures above 20 TeV~\cite{Ellis:2020nnr} which perfectly fits the range of parameters within the reach of LISA (see figure~\ref{fig_gamma_uncertainty}). In the case of the model with a new SU(2) group, it has been shown in ref.~\cite{Kierkla:2022odc} that the case of non-instantaneous reheating ($\gf<H_*$) cannot be realised since in this region of parameter space the phase transition does not end successfully with percolation of bubbles. This conclusion could be modified by updates in the percolation criterion, see e.g.~ref.~\cite{Lewicki:2024sfw}. Then, according to ref.~\cite{Kierkla:2022odc} (see also ref.~\cite{Kierkla:2023von} for an NLO update), in the region of small $\gf$ the expected values would be around $T_{\rm reh}\approx 10^5~\rm{GeV}$,\footnote{For exact values more detailed modelling of inefficient reheating would be needed which was not done in ref.~\cite{Kierkla:2022odc}} and $\beta/H_*\approx 10$, which overlaps with the range where we predict $\gf$ could be probed. Therefore, we can take the values of $T_{\rm reh}$ and $\beta/H_*$ as indicative for a class of similar models.  This shows that GWs from supercooled phase transitions could directly probe underlying fundamental physics models.

%~~~~~~~~~~~~~~~~~~~~~~~~~~~~~~~~~~~~~~~~~~~~~~~~~~~~~~~~~~~~~
\section{Conclusions}\label{sec:conclusions}
%~~~~~~~~~~~~~~~~~~~~~~~~~~~~~~~~~~~~~~~~~~~~~~~~~~~~~~~~~~~~~

In this work, we presented a projection for the reconstruction of parameters describing the GW spectra in supercooled cosmological phase transitions using the Fisher matrix formalism, for two forthcoming GW detectors, LISA and ET. We used both the simplified spectra in the geometrical parameterisation (peak amplitude, $\Omega_p$ and frequency, $f_p$ as well as $d$ -- the slope of the spectrum for frequencies corresponding to super-Hubble scales) and the spectra parameterised by the thermodynamical quantities (inverse time scale of the transition $\beta/H_*$, maximal temperature attained during reheating, $T_{\rm{ max}}$ and the decay rate of the scalar field, $\gf$). In the latter case, the dilution of the signal due to an early matter-domination period is properly taken into account. 

We showed that for strong signals associated with supercooled PTs one can probe the parameters of the spectra with a very good precision. In a wide range of parameter space admitting supercooling, individual parameters (both geometrical and thermodynamical) could in principle be reconstructed with accuracy at percent level. This accuracy, however, is worsened when degeneracies among the thermodynamical parameters are taken into account.

If the transition is strong enough we will gain access to the low-frequency slope going beyond the frequencies corresponding to the horizon size at the time of the transition. tilt of the slope beyond that scale encodes information on the expansion rate following the transition and allows us to probe the process of reheating. As our two benchmarks, we used the standard radiation domination following the transition ($d=3$) and a period of matter domination ($d=1$) corresponding to inefficient reheating where the field oscillates for a significant amount of time around the true vacuum before decaying into SM particles. We found that in a small yet viable parameter space, the accuracy of reconstructing the tilt $d$ is enough to distinguish between early matter domination and standard radiation domination. Thus, it is possible that GW signals from strongly supercooled phase transitions will be used to probe the expansion history of the universe.

Typically the thermodynamical parameters defining the spectrum are degenerate and thus many points in the parameter space of a given model can result in the same GW spectra. The reheating history is different in that regard as it has a direct impact on the shape of the spectrum. We described the inefficient reheating in terms of the underlying particle physics model using the decay rate of the scalar $\gf$. The ratio of this parameter and the Hubble constant determines the length of the matter domination period. As a result, it controls the range of frequencies where the characteristic plateau in the spectrum is produced. We show that this parameter can be measured provided the amplitude of the spectrum is large enough. Whenever this is possible we would be able to gain direct information about the underlying fundamental physics model from the GW spectrum. Interestingly, the phase of early matter-domination can appear if the new scalar undergoing the phase transition has a small decay rate, in particular, it is very weakly coupled to the SM Higgs boson. In this way, cosmological GW would constitute a way to probe couplings probably inaccessible at colliders, proving synergy between these two approaches to probe fundamental interactions.

\section*{Acknowledgments}
\newcommand{\kT}[1]{{\fontencoding{T1}\selectfont\k{#1}}}
We would like to thank Jakub Br\kT{e}czewski, Adam Gomu{\l}ka, Marek Soko{\l}owski and Wojciech Olejko for verification of an early version of the code.
At the early stages, the work of B\'S and AG was supported by the National Science Centre, Poland, through the SONATA
project number 2018/31/D/ST2/03302. Later, the work of B\'S was funded by the National Science Centre, Poland,
through the OPUS project number 2023/49/B/ST2/02782.
This work was supported by the Polish National Agency for Academic Exchange within the Polish Returns Programme under agreement PPN/PPO/2020/1/00013/U/00001 and the National Science Center, Poland, through Sonata Bis grant 2023/50/E/ST2/00177.
This research was funded in part by the National Science Centre, Poland, through the OPUS project number 2023/49/B/ST2/02782. For the purpose of Open Access, the authors have applied a CC-BY public copyright licence to any Author Accepted Manuscript version arising from this submission.
\bigskip

\noindent \textbf{Data Availability Statement.} The data used to prepare the figures is available at ref.~\cite{data_source}. 
%The data used to prepare the plots is available at \cite{data_source}.
\bibliographystyle{JHEP.bst}
\bibliography{bib_final.bib}
\end{document}